\documentclass[a4paper,fleqn,usenatbib]{mnras}
\usepackage{newtxtext,newtxmath}
\usepackage[T1]{fontenc}
\usepackage{ae,aecompl}
           
\usepackage{graphicx} 
\usepackage{natbib}   
\usepackage{ulem}

\newcommand{\HI}{H\,{\sc i}}

\newcommand{\kms}{~km\,s$^{-1}$}
\newcommand{\kkms}{km\,s$^{-1}$}

\newcommand{\vopt}{$v_{\rm opt}$}
\newcommand{\vHI}{$v_{\rm HI}$}

\newcommand{\vsys}{$v_{\rm sys}$}

\newcommand{\FHI}{$F_{\rm HI}$}

\newcommand{\NHI}{$N_{\rm HI}$}
\newcommand{\MHI}{$M_{\rm HI}$}

\newcommand{\Msun}{~M$_{\odot}$}

\title[HI debris in the IC\,1459 galaxy group]
      {\HI\ debris in the IC~1459 galaxy group\thanks{The radio data presented 
  here were obtained with the Australia Telescope Compact Array (ATCA) -through the Australia Telescope Online Archive- and the 
  64-m Parkes telescope, which are funded by the Commonwealth of Australia for 
  operation as a National Facility managed by CSIRO.} }

\author[J.~Saponara et al.]
    {Juliana Saponara,$^{1,2}$\thanks{E-mail: jsaponara@iar.unlp.edu.ar}
     B\"arbel S. Koribalski,$^3$
     Paula Benaglia,$^{1,2}$ and \newauthor
     Manuel Fern\'andez L\'opez$^{1}$ \\ \\
     $^{1}$Instituto Argentino de Radioastronom\'{\i}a, (CONICET; CICPBA), C.C. No. 5, 1894,Villa Elisa, Argentina \\
     $^{2}$Facultad de Ciencias Astron\'omicas y Geof\'{\i}sicas, Universidad Nacional 
           de La Plata, Paseo del Bosque s/n, 1900 La Plata, Argentina \\
     $^{3}$CSIRO Astronomy \& Space Science, 
           Australia Telescope National Facility, 
           P.O. Box 76, Epping, NSW 1710, Australia
}

\date{Received date; accepted date}

\pubyear{2016}

\begin{document}
\maketitle

\begin{abstract}
We present \HI\ synthesis imaging of the giant elliptical galaxy IC\,1459 and
its surroundings with the Australia Telescope Compact Array (ATCA). Our search
for extended \HI\ emission revealed a large complex of \HI\ clouds near 
IC\,1459, likely the debris from tidal interactions with neighbouring galaxies. 
The total \HI\ mass ($\sim 10^9$\Msun) in the detected clouds spans 250 kpc 
from the north-east of the gas-rich spiral NGC~7418A to the south-east of 
IC\,1459. The extent and mass of the \HI\ debris, which shows rather irregular 
morphology and kinematics, are similar to those in other nearby groups. 
Together with \HI\ clouds recently detected near two other IC\,1459 group 
members, namely IC\,5270 and NGC~7418, using Phased-Array Feeds (PAFs) on the 
Australian Square Kilometer Array Pathfinder (ASKAP), the detected debris 
make up a significant fraction of the group's intergalactic medium.
\end{abstract}


\section{Introduction} 
Gravitational interactions between galaxies and their environment play an 
important and on-going role in galaxy evolution, affecting their local and 
global properties such as gas content, kinematics and star formation rate. 
Physical processes like gas accretion, tidal effects and ram pressure 
stripping shape the outskirts of galaxy discs, most noticeably their \HI\ 
extent and morphology \citep{2011MNRAS.410.2217W}. Furthermore, the kinematics 
of the \HI\ gas associated with galaxies and/or tidal debris can highlight 
the past orbits of their interactions.
 Though galaxy transformations can happen in lower density environments  \citep{2011MNRAS.416.1680C}, they occur more often in dense groups and clusters. Substantial \HI\ debris are typically 
detected near massive, often early-type galaxies \citep{1989ApJ...343...94S, 
2001ApJ...555..232R, 2005MNRAS.357L..21B, 2005MNRAS.363L..21B, 
2009AJ....138.1741C,2010AJ....139..102E, 2010A&A...515A..67S}. In particular,
we highlight the ATLAS-3D \HI\ results for a large sample of nearby early-type
galaxies, which show a 40\% \HI\ detection rate for the
non-cluster targets \citep{2012MNRAS.422.1835S}. While most of the detections
consist of \HI\ discs or rings, \HI\ debris in form of filaments and clouds 
are typical in rich groups.

The high density of galaxies in clusters together with their high velocity 
dispersion typically lead to significant gas stripping, particularly from the 
outer discs of gas-rich spirals, resulting in mainly \HI-deficient galaxies 
near the cluster centre \citep{2007ApJ...659L.115C, 2009AJ....138.1741C}. The 
observed \HI\ filaments in, for example, Virgo and Ursa Major 
\citep{2005A&A...437L..19O, 2006PASP..118..517B, 2001ASPC..240..867V,
2013MNRAS.428.1790W} hint at the disruptive processes under way, while 
providing fuel for the formation of new dwarf galaxies from the tidal debris 
\citep{2004A&A...427..803D, 2016MNRAS.460.2945L}.

\begin{table*} 
\caption{Optical properties of IC\,1459 group members in the observed area.}
\centering
\begin{tabular}{lccccccccc}
\hline
\hline
Galaxy name& type & centre position&\vopt&$m_{\rm B}$&$D_{\rm B25}$& $i$&$PA$
  & offset from IC\,1459 \\
           &      & $\alpha,\delta$(J2000)&[\kkms]&[mag]&[arcmin]&[deg]&[deg]
  & [arcmin] \\
     \hline
IC~5269B & SBcd  & 22:56:36.5, --36:14:57& 1659 & 13.2 & 4.1 & 78 & 96 & 14.5\\
IC~5269A & SBm   & 22:55:55.7, --36:20:52& 2870 & 13.9 & 1.3 & 41 & 35 & 16.5\\
IC~1459  &  E3   & 22:57:11.5, --36:27:44& 1691 & 11.0 & 4.9 & 42 & 47 &  ---\\
dE1      & dE0/Im& 22:57:07.6, --36:37:15& ---  & 15.5 & 0.04& ---& ---&  9.5\\
dE2      & dE0,N & 22:57:10.6, --36:40:09& 1838 & 17.3 & 0.32& ---& ---& 12.4\\
IC~5264  & Sab   & 22:56:52.6, --36:33:15& 2043 & 13.7 & 2.5 & 77 & 82 &  6.5\\
NGC~7418A& SAd   & 22:56:41.8, --36:46:21& 2102 & 13.8 & 1.8 & 56 & 83 & 19.5\\
\hline
\end{tabular}
\flushleft Notes: Cols.~(2--8) give the galaxy type, its centre position 
  [$^{\rm h, m, s}$, \degr\,\arcmin\,\arcsec], centre velocity (\vopt), blue 
  magnitude ($m_{\rm B}$), diameter ($D_{\rm B25}$) measured at the 25th blue 
  magnitude and the corresponding disc inclination ($i$) and position angle 
  ($PA$). --- References: ESO LV \cite{1989spce.book.....L}, \cite{RC3}, 
  \cite{1991ApJS...75..935D}, and \cite{1994AJ....108.1209V}.
\label{tab:optical}
\end{table*}

\begin{table*} 
\caption{ATCA \HI\ observations of the IC\,1459 galaxy group.}
\begin{tabular}{ccccc}
\hline
\hline
ATCA configuration     & H214     & 375      & 750A     & 6A       \\
\hline
Project                & C1154    & C530     & C1027    & C689     \\
Date                   & 04-03-05 & 19-09-96 & 08-01-02 & 03-02-98 \\
                       &          & 23-09-97 &          & 08-02-98 \\
                       &          &          &          & 09-02-98 \\
Time on-source [min.]  &   87     & 396      & 554.3    & 163.5    \\
                       &          & 651      &          &  40      \\
                       &          &          &          &  80      \\
Primary calibrator     & \multicolumn{4}{c}{PKS~1934--638 (14.95 Jy)} \\
Phase calibrator       & 0008--421& 0008--421& 2259--375& 2259--375 \\
Centre frequency [MHz] & 1412     & 1412     &  1413    &  1413   \\
Bandwidth [MHz]        &    8     &   16     &     8    &     8   \\
No. of channels        &  512     &  512     &  1024    &   512   \\
Channel width [\kkms]  &  3.3     &  6.6     &   1.6    &   3.3   \\
Velocity resolution [\kkms]  & 4  &  8       &   2      &     4   \\
\hline
\end{tabular}
\label{tab:obs}
\end{table*}

\begin{table*}  
\caption{HIPASS properties of IC\,1459 group members in the observed area.}
\begin{tabular}{llcccccc}
\hline
\hline
Galaxy name & HIPASS name &\vsys&$w_{\rm 50}$&$w_{\rm 20}$ & \FHI  & Reference 
   & offset from \\
            &             & [\kkms] & [\kkms] & [\kkms] & [Jy\kms] & 
   & optical position \\
\hline
IC\,5269B    & HIPASS J2256--36b    & $1667 \pm 4$ & 226 & 242 & $21.5 \pm 3.3$ 
   & (1) & 1\farcm5 \\
IC\,1459 debris&HIPASS J2257--36$^*$& $1802 \pm12$ &  80 & 111 & 3.0 
   & (2) & 3\farcm5 \\
IC\,5269A    & HIPASS J2255--36     & $2872 \pm 8$ & 124 & 147 & 6.0 
   & (2) & 1\farcm3 \\
IC\,5264     & ~~~~~~ ---           & $\sim$1950   &     & $\sim$300 & 
   & here \\ 
NGC~7418A    & HIPASS J2256--36a    & $2115 \pm 5$ & 173 & 209 & $23.5 \pm 3.3$
   & (1) & 1\farcm0 \\
\hline
\end{tabular}
\flushleft Notes: Cols.~(3--6) give the \HI\ systemic velocity (\vsys), 50\% 
  and 20\% velocity width ($w_{\rm 50}$, $w_{\rm 20}$) and integrated \HI\ 
  flux density (\FHI). --- $^*$~HIPASS J2257--36 encompasses some of the \HI\ 
  debris detected near IC\,1459.  --- References: 
       (1) \citet{2004AJ....128...16K} (HIPASS BGC); 
       (2) \citet{2004MNRAS.350.1195M} (HICAT).
\label{tab:hipass}
\end{table*}

The galaxy group environment is less dense, and interactions occur at much 
lower speeds than in clusters, allowing us to study tidal interactions, ram 
pressure stripping, minor mergers and gas accretion. 
Many detailed \HI\ 
studies, ranging from galaxies in close pairs \citep[e.g.,][]{2001MNRAS.326..578G, 2004MNRAS.348.1255K, 2009MNRAS.400.1749K, 2015A&A...584A.114S}, compact 
groups \citep{2003MNRAS.342..939G, 2003MNRAS.339.1203K, 2012MNRAS.422.1835S}, 
Hickson compact groups \citep{1982ApJ...255..382H,2001A&A...377..812V}, loose 
groups \citep{2009MNRAS.400.1962K, 2011ApJS..197...28P} and clusters 
\citep{2009AJ....138.1741C} have been carried out. The stripped gas may be 
found as tidal tails, filaments, and plumes near the interacting galaxies, 
e.g. in the Abell\,1367 cluster \citep{2010MNRAS.403.1175S} and nearby groups 
\citep[e.g.,][]{1996ApJ...467..241H, 2003MNRAS.339.1203K, 2004MNRAS.348.1255K, 
2010AJ....139..102E}.
These extended \HI\ structures can be difficult to detect, requiring high 
sensitivity single-dish observations to measure the low-surface brightness
emission and long interferometric observations with emphasis on short 
baselines such as not to resolve out the diffuse \HI\ emission.

The IC\,1459 loose galaxy group provides a particularly interesting environment
to study the displacement of neutral hydrogen (\HI) due to tidal interactions 
and ram pressure stripping, since IC\,1459 is a giant elliptical galaxy 
surrounded by at least 15 neighbours \citep{1990ApJ...352..532W, 
2009MNRAS.400.1962K}. This group was recently observed by 
\citet{2015MNRAS.452.2680S} using the first six PAF-equipped antennas of the 
Australian Square Kilometre Array Pathfinder\footnote{The wide field-of-view 
(30 sq degrees) delivered by the ASKAP Phased-Array Feeds (PAFs) with all 36 
beams is particularly suitable to study the \HI\ emission in nearby galaxy 
groups and clusters.} \citep[ASKAP]{2014PASA...31...41H}. The wide-field 
ASKAP \HI\ data, initially obtained with nine $\sim$1 deg PAF beams, resulted 
in the detection of 11 galaxies and three \HI\ clouds, two in the proximity 
of the edge-on spiral galaxy IC\,5270 and one close to the spiral galaxy 
NGC~7418 \citep{2015MNRAS.452.2680S}. Here we explore the area around IC\,1459 
using single-beam Australia Telescope Compact Array (ATCA) archival \HI\ data.

\subsection{The IC\,1459 galaxy group} 
IC\,1459 is an early-type galaxy located in the centre of a loose group at 
an approximate distance of 29~Mpc \citep{2001MNRAS.327.1004B, 2001ApJ...546..681T}; here we adopt this distance for the whole group \citep[see also][]{2015MNRAS.452.2680S}. The IC\,1459 galaxy group contains ten bright galaxies, mostly of late-type morphology, \citep{2006MNRAS.370.1223B, 2009MNRAS.400.1962K}, which form part of the larger-scale Grus cluster \citep{1981MNRAS.195P...1A, 1982ApJ...257..423H}.
In addition, two dwarf elliptical galaxies (here denoted dE1 and dE2) have been
catalogued by \citet{1994AJ....108.1209V} as group members, both located 
south-east of IC\,5264. While the nucleated dwarf dE2 is clearly visible in 
second-generation Digitized Sky Survey (DSS2) images, its low-surface 
brightness neighbour dE1 is hard to see.
Table~\ref{tab:optical} summarises the optical properties of IC\,1459 group members in our 
study while Table~\ref{tab:obs} lists the \HI\ properties from the \HI\ Parkes All Sky 
Survey \cite[HIPASS]{2004AJ....128...16K}. 

Deep optical images of the IC\,1459 group by \citet{malin1985} and 
\citet{1995AJ....109.1576F} reveal very faint outer arms, plumes and shells 
in the outskirts of IC\,1459's stellar body, indicative of accretion and 
merger activity. The faintest shells appear to encompass the neighbouring 
spiral galaxy IC\,5264. The average distance of the largest shell from the 
centre of IC\,1459 is 3\farcm8. \citet{1995AJ....109.1576F} find the shells, 
which are oriented along the major axis of the galaxy, to have red colours 
similar to the main galaxy.

A low-sensitivity VLA \HI\ map of IC\,1459 and its surroundings is presented 
by \citet{1990ApJ...352..532W}, who detect three group members (IC\,5269B, 
NGC~7418A and IC\,5264) and give an upper limit of 3 Jy\kms\ for the \HI\ flux 
in the close vicinity of IC\,1459 (ie. \MHI\ $< 6 \times 10^8$\Msun). Their 
\HI\ channel maps show hints of \HI\ debris east of IC\,5264 at $\sim$1800\kms.
A preliminary ATCA \HI\ map of the same area is shown by 
\citet{1999ASPC..163...72O} who clearly detect extended debris around IC\,1459
and briefly discuss their irregular morphology, suggesting that \HI\ stripping 
from the gas-rich group members through interactions with IC\,1459 is their 
likely origin.  

ROSAT X-ray observations by \citet{2004MNRAS.350.1511O} reveal very low-level 
extended emission near IC\,1459 and a total luminosity for the group of 
$L_{\rm X}$ ($r_{\rm 500}$) = 41.51 erg\,s$^{-1}$. \citet{2009MNRAS.400.1962K} 
conducted Parkes \HI\ observations for 16 galaxy groups and compare their 
\HI\ and X-ray properties. For the IC\,1459 group they derive a total \HI\ 
mass at least of $2.5 \times 10^{10}$\Msun. Based on IC\,1459's dynamically 
inferred black hole mass of $3 - 12 \times 10^8$\Msun, 
\citet{2000AJ....120.1221V} estimate a total galaxy mass of $\sim$3 $\times
10^{11}$\Msun\ (all quoted mass estimates are adjusted to our adopted 
distance). Using dynamic modelling of the HST data they measure a systemic 
velocity of \vsys\ = $1783 \pm 10$\kms, of the ionised gas in IC\,1459,
$\sim$90\kms\ higher than IC\,1459's optical velocity of $1691 \pm 18$\kms\ 
\citep{RC3}.
A narrow (35\kms) $^{12}$CO(1--0) line emission 
towards IC\,1459 at \vsys\ = 1782\kms\ was detected by 
\citet{2001A&A...374..421B}. They give a molecular mass limit of $2 \times 
10^7$\Msun\ and suggest the emission originates from a giant molecular 
association, possibly a merger residual. 

The IC\,1459 galaxy itself has a number of peculiar features, hinting at 
disturbances due to interactions and/or merging: twisted isophotes first
noticed by \citet{1979ApJ...227...56W} and \citet{1989ApJ...344..613F}, a 
fast counter-rotating stellar core \citep{1988ApJ...327L..55F}, faint stellar 
shells \citep{malin1985} and a large H $+$ N\,[{\sc iii}] emission-line disc 
showing a weak spiral structure \citep{1990A&A...228L...9G}. Furthermore, 
IC\,1459 hosts a bright radio source, PKS~2254--367, characterised by two 
symmetric radio jets, whose activity may have been triggered by the same 
events that gave the galaxy its peculiar morphology and kinematics. The jets 
extend to only 8 pc, well within the galaxy core, indicating the AGN is very 
young \citep{2015MNRAS.448..252T}. \\

\begin{table*} 
\caption{ATCA \HI\ properties of individual clouds near IC\,1459.}
\begin{tabular}{lcccccccc} 
\hline
\hline
\HI\ clouds & centre position & velocity range & \multicolumn{2}{c}{\FHI} 
   & \multicolumn{2}{c}{\MHI} & nearest galaxy \\
     & $\alpha,\delta$ (J2000)& [\kkms] & \multicolumn{2}{c}{[Jy\kms]} 
   & \multicolumn{2}{c}{[$10^8$\Msun]} & \\
\hline
 C1 & 22:57:44, --36:33:13 & 1780 -- 1844 & 0.81 $\pm$ 0.16 & 0.55 $\pm$ 0.02 & 1.61 & 1.09  & IC\,5264 \\
 C2 & 22:57:42, --36:38:08 & 1752 -- 1792 & 0.35 $\pm$ 0.10 & 0.18 $\pm$ 0.02 & 0.70 & 0.35  & NGC~7418A \\
 C3 & 22:57:27, --36:37:13 & 1776 -- 1800 & 0.18 $\pm$ 0.05 &  ---            & 0.36 &  ---  & NGC~7418A \\
 C4 & 22:57:17, --36:34:09 & 1744 -- 1808 & 0.83 $\pm$ 0.10 & 0.15 $\pm$ 0.01 & 1.66 & 0.29  & IC\,5264\\
 C5 & 22:57:13, --36:30:10 & 1759 -- 1828 & 0.43 $\pm$ 0.10 & 0.14 $\pm$ 0.02 & 0.85 & 0.27  & IC\,1459\\
\hline
 C6 & 22:57:04, --36:33:35 & 1480 -- 1512 & 0.22 $\pm$ 0.05 & ---             & 0.5 &--      & IC\,5264 \\
 C7 & 22:56:54, --36:42:09 & 1680 -- 1736 &  0.7 $\pm$ 0.1  & ---             & 1.5 &---     & NGC~7418A \\
\hline 
\label{tab:HIclouds}
\end{tabular}
\flushleft Notes: Cols.~(2--6) give the centre position [$^{\rm h, m, s}$, 
  \degr\,\arcmin\,\arcsec], \HI\ velocity range, and two values for the \HI\ 
  flux density (\FHI) and \HI\ mass (\MHI) of the identified \HI\ clouds,
  as obtained from the low- and high-resolution maps, respectively (see 
  Section~2).  
\end{table*}

\begin{figure*}  
 \includegraphics[]{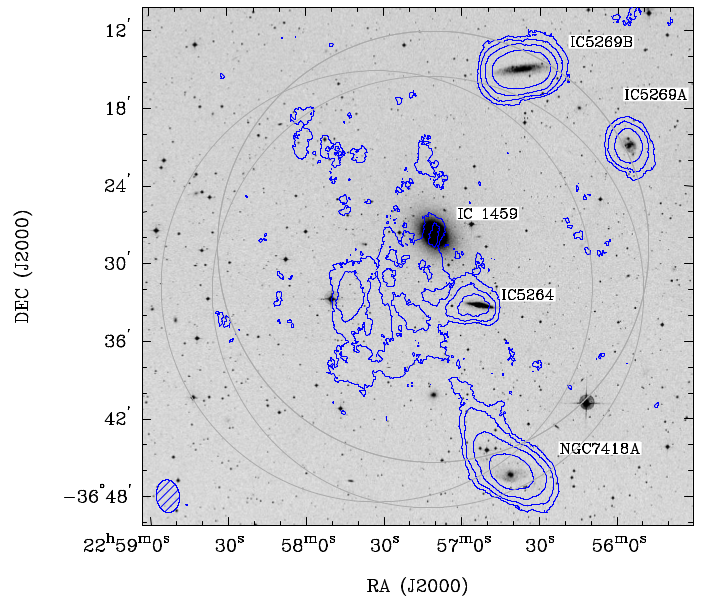} 
\caption{ATCA \HI\ distribution of the IC~1459 group in the velocity range 
    from 1200 to 2950\kms, overlaid onto a DSS2 $B$-band image. Large amounts
    of \HI\ debris are detected around the giant elliptical galaxy IC\,1459 
    as well as \HI\ emission in four galaxies (all known group members are 
    labelled). The contour levels are 0.25, 0.5, 1, and 2 Jy\,beam$^{-1}$\kms. 
    The synthesised beam ($152\arcsec \times 107\arcsec$) is displayed in the 
    bottom left corner; faint grey circles mark the three different locations
    of the ATCA primary beam. This image is not primary beam corrected.}
\label{fig:ATCAfield}
\end{figure*}

In this paper, we present ATCA \HI\ data that allow us to study the diffuse 
\HI\ emission between IC\,1459 and its nearest neighbours. In Section~2 we 
describe the ATCA observations and data reduction, followed by our results 
in Section~3 and a discussion in Section~4. A summary and an outlook are given 
in Section~5.

\section{Observations \& data reduction} 

\HI\ line and 20 cm continuum observations of IC\,1459 were obtained with the 
Australia Telescope Compact Array (ATCA) in the H214, 375, 750A, and 6A 
configurations between September 1996 and March 2005. The data were
downloaded from the Australia Telescope Online Archive\footnote{http://atoa.atnf.csiro.au/} (ATOA).
The observing parameters are summarised in Table~\ref{tab:obs};
the total time spent on the target field was 32.8 hours.

Data reduction and analysis were performed with the {\sc miriad} software 
package \citep{1995ASPC...77..433S} using standard procedures. In addition, 
we extensively used {\sc kvis}, part of the {\sc karma} package 
\citep{1996ASPC..101...80G}, for the visualisation of our data cubes and 
multi-wavelength images. We calibrated each data set separately, using 
PKS~1934--638 as primary flux and bandpass calibrator except for the 375 
configuration, where PKS~2259--375 was used for the bandpass calibration. 
The latter data set is somewhat affected by solar interference.

PKS~2259--375 and PKS~0008--421 served as the phase calibrators. Using a flux 
density of 14.95 Jy for PKS~1934--638, we obtained 2.6 Jy for PKS~2259--375 
and 4.3 Jy for PKS~0008--421. Continuum subtraction was made using {\sc UVLIN};
when required options {\em sun} and {\em twofit} were used to subtract 
continuum emission from the Sun and radio sources in our field. With a 
bandwidth of at least 8~MHz (see Table~\ref{tab:obs}), we were able to select line-free 
channel on either side of the detected \HI\ emission within the IC\,1459 
galaxy group. The \HI\ spectral line data were Fourier-transformed using 
`natural' and `robust' weighting. The longest baselines to the distant 
antenna six (CA06) were excluded when making the low-resolution \HI\ cubes 
to enhance sensitivity for diffuse, extended \HI\ emission. As the four data 
sets have three different pointing centres, marked in Fig.~\ref{fig:ATCAfield},
a mosaic image was produced, then CLEANed and restored. The synthesised beam 
in the low-resolution \HI\ maps is $152\arcsec \times 107\arcsec$, and we 
measure an r.m.s. noise of $\sim$1.7 mJy\,beam$^{-1}$ per 8\kms\ channel in 
the centre of the field. The 5$\sigma$ 
$N_{\rm HI}$ and \MHI\ limits over three channels are
$\sim 6 \times 10^{17}$ cm$^{-2}$ and $\sim$4 $\times 10^7$\Msun. 
The synthesised beam in the high-resolution \HI\ 
maps is $30\arcsec \times 30\arcsec$. The corresponding r.m.s. noise, the 5$\sigma$ 
$N_{\rm HI}$ and \MHI\ limits over three channels of the high resolution cube, used to identify high-density \HI\ clumps,
are $\sim0.8$ mJy beam$^{-1}$ per 8 km s$^{-1}$, $\sim 5 \times 10^{18}$cm${-2}$ and $\sim0.2 \times 10^7$\Msun.


\section{Results} 

The ATCA \HI\ distribution within the IC\,1459 galaxy group for the observed 
region and velocity range is shown in Fig.~\ref{fig:ATCAfield}. We detect \HI\ 
emission in four spiral galaxies; these are (from north to south) IC\,5269B, 
IC\,5269A, IC\,5264 and NGC~7418A. Their published optical and HIPASS 
\citep{2004AJ....128...16K} properties are listed in Tables~\ref{tab:optical} 
\& \ref{tab:hipass}, respectively. We also find a complex of \HI\ clouds near 
the elliptical galaxy IC\,1459 as well as individual \HI\ clouds north-east 
of NGC~7418A and east of IC\,5264.
 Even though the IC\,1459 group contains a diffuse hot/warm intra-group medium and probably
ram pressure stripping played a role, we found relevant evidende pointing at tidal interactions 
beween group members. The cloud properties, as measured 
from the presented ATCA \HI\ data, are summarised in Table~\ref{tab:HIclouds}. 
These \HI\ clouds are in addition to those discovered in ASKAP \HI\ data of 
the IC\,1459 galaxy group by \citet{2015MNRAS.452.2680S} near the spiral 
galaxies IC\,5270 and NGC~7418, both of which lie outside the area observed 
here. The lowest contour in Fig.~\ref{fig:ATCAfield} corresponds to an \HI\ 
column density of \NHI\ = $1.7 \times 10^{19}$\,cm$^{-2}$ (assuming the \HI\ 
gas fills the beam). 

%


\subsection{\HI\ debris near IC\,1459} 

\begin{figure*} 
  \includegraphics[]{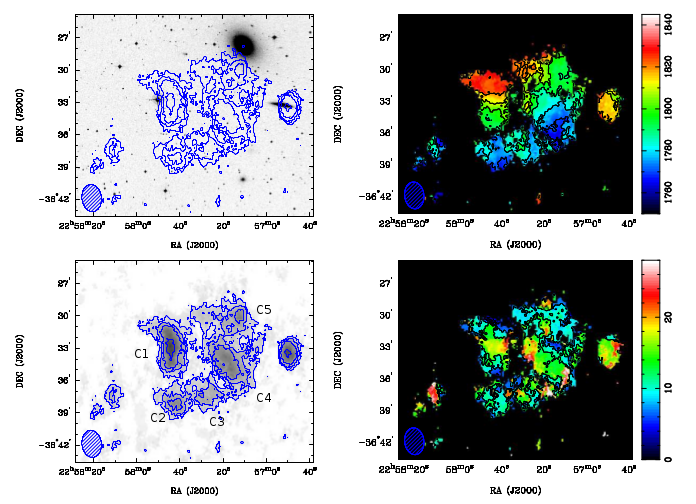} 
\caption{ATCA \HI\ moment maps of tidal debris south-east of the giant 
  elliptical galaxy IC\,1459 in the velocity range from 1750 to 1840\kms. Some \HI\ emission associated with the small edge-on galaxy IC\,5264 is also visible in this velocity range. {\bf (Top Left:)} \HI\ distribution (contours) overlaid onto the DSS2 $B$-band image; the contour levels are 0.7, 1.4, 2.1 and 3.5 $\times 10^{19}$ cm$^{-2}$.
{\bf (Bottom Left:)} \HI\ distribution (same contours); {\bf (Top Right:)} Mean \HI\ velocity field; the contour levels range from 1760 to 1820\kms\ in steps of 10\kms. {\bf (Bottom Right:)} Mean \HI\ velocity dispersion; the contour levels are 4, 8, and 12\kms. The synthesised beam ($151\farcs9 \times 108\farcs7$) is shown in the bottom left corner of each panel.}
\label{fig:HIclouds}
\end{figure*}

\begin{figure*} 
%
%
 \includegraphics[width=1\textwidth]{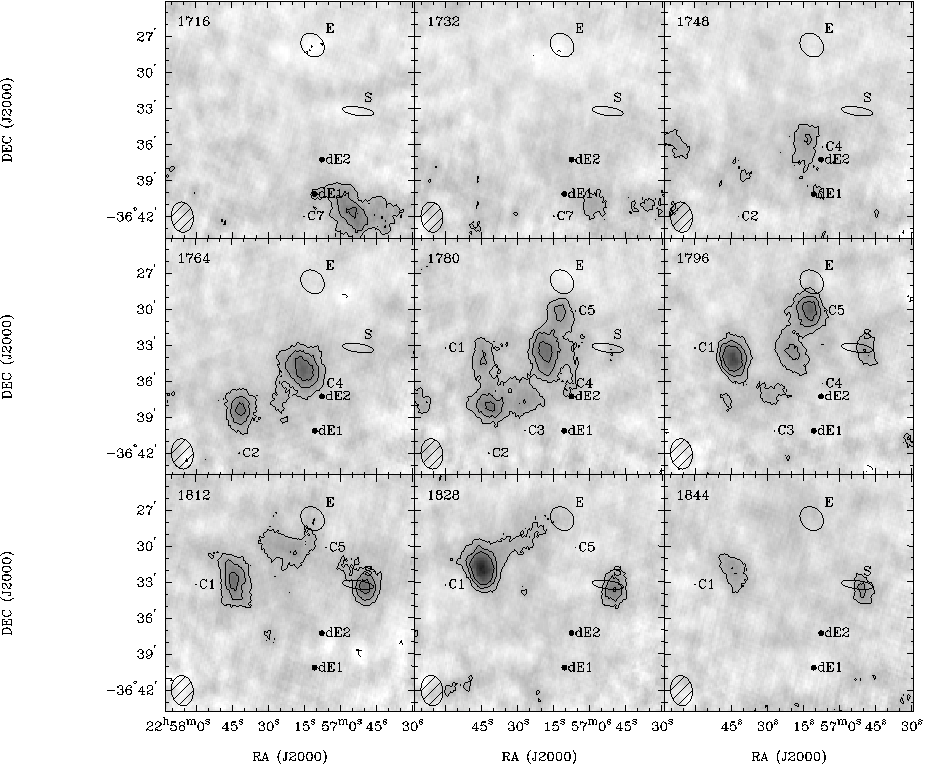}
\caption{ATCA \HI\ channel maps of the cloud complex around IC\,1459. 
   The channel velocity is shown in the top left corner (in \kms) and the 
   synthesised beam in the bottom left corner of each panel. The positions 
   of the galaxies IC\,1459 (E) and IC\,5264 (S) as well as dE1, dE2 and the clouds are 
   marked. The contour levels are 5.1, 8.5, and 11.9 mJy beam$^{-1}$.}
\label{fig:channels} 
\end{figure*}

The \HI\ debris in the surroundings of IC\,1459 are wide-spread (see Fig.~\ref{fig:ATCAfield})
and difficult to fully map with an interferometer due to their low surface 
brightness and extended nature. In Fig.~\ref{fig:HIclouds} we focus on the 
brightest \HI\ cloud complex, located south-east of IC\,1459 and IC\,5264, 
spanning a narrow velocity range from $\sim$1760 to 1850\kms. ATCA \HI\ channel
maps of these \HI\ clouds, smoothed to 16\kms\ resolution, are shown in 
Fig.~\ref{fig:channels}.
The \HI\ debris complex shown in Fig.~\ref{fig:HIclouds} spans $\sim$10\arcmin\ (or 85 kpc at 
the assumed distance of 29 Mpc). The mean \HI\ velocity field of the debris 
shows a weak velocity gradient, generally decreasing from north to south. The 
mean \HI\ velocity dispersion is low, compared to that found in galaxies, and 
consistent with a cloud complex. The full extent of the \HI\ debris shown in 
Fig.~\ref{fig:ATCAfield} appears to be at least $\sim$3 times larger than the
cloud complex, spanning over $\sim$250~kpc from the eastern tip of NGC~7418A 
to the \HI\ clouds north-east of IC\,1459.  \\ 

The \HI\ debris near IC\,1459 were first detected by \cite{1999ASPC..163...72O}
with the ATCA and later independently discovered in HIPASS, catalogued as 
HIPASS J2257--36 \citep{2004MNRAS.350.1195M}. The HIPASS properties are given
in Table~\ref{tab:hipass}. We note that the fitted HIPASS position, $\alpha,\delta$(J2000) 
= $22^{\rm h}\,57^{\rm m}\,12.6^{\rm s}$, --36\degr\,31\arcmin\,14\arcsec, 
is $\sim$3\farcm5 offset from the optical centre of IC\,1459 to the south-east 
and 4\farcm4 from IC\,5264 to the east, in agreement with the \HI\ debris
complex identified here (see Fig.~\ref{fig:HIclouds}). The centre velocity of HIPASS J2257--36 
($1802 \pm 12$\kms) and its velocity widths (see Table~\ref{tab:hipass}) also match those
obtained here. Using our ATCA data we measure \FHI\ = 2.6 Jy\kms\ for the \HI\ 
debris complex shown in Fig.~\ref{fig:HIclouds}, in agreement with HIPASS, and derive an \HI\ 
mass of $5.2 \times 10^8$\Msun. For IC\,5264, located just west of the \HI\
debris complex, we measure \FHI\ = $0.5 \pm 0.2$ Jy\kms.\\



We identified five \HI\ clouds (C1 -- C5) within the \HI\ debris complex shown 
in Fig.~\ref{fig:HIclouds}; their ATCA \HI\ properties are given in Table~\ref{tab:HIclouds}.
No optical or GALEX counterparts associated with these clouds are detected in 
currently available survey data. Our high-resolution (30\arcsec) ATCA \HI\ 
maps (not shown here) reveal the densest clumps within these clouds. Four of the 
five \HI\ clouds show compact clumps with \HI\ masses of 0.1--1.1 $\times 
10^8$\Msun. Two additional \HI\ clouds, C6 and C7, can be disentangled from 
their nearest galaxies due to their distinct radial velocities, $\sim$1480\kms\ 
and 1704\kms, respectively. 

The detected \HI\ debris near IC\,1459 together with \HI\ clouds near IC\,5270 
($\sim$2100\kms) and NGC~7418 ($\sim$1450\kms), as identified by 
\citet{2015MNRAS.452.2680S}, make up a significant fraction of the groups' 
cold interstellar medium. 
 The total amount of \HI\ mass contained in clouds (7.2$\times 10^8$~\Msun) is almost 6 times
the \HI\ content of IC\,5264 (1.2$\times 10^8$~\Msun). According to \citet{2014MNRAS.444..667D}
scaling relations, IC\,5264 is an \HI\ deficient galaxy. The scenario resembles the first stages of a process that yields to galaxies following the morphology-density relation and becoming \HI\ deficient.

Searching for \HI\ absorption against the bright radio continuum source
PKS~2254--367 ($\sim$1.1~Jy at 21-cm) at the centre of the elliptical galaxy 
IC\,1459 we find a weak line (2.5 sigma level) at $\sim$1714\kms\ with a velocity width of 
$\sim$30\kms. The closest \HI\ clump to IC\,1459 within the debris complex 
appears to be C5 (labelled in Fig.~\ref{fig:HIclouds}), most prominent at velocities of 
1780--1800\kms\ (see Fig.~\ref{fig:channels}). The CO(1--0) emission detected at 1782\kms\ 
by \citet{2001A&A...374..421B} also falls into this range.

\begin{figure*}  
 \includegraphics[width=0.8\textwidth]{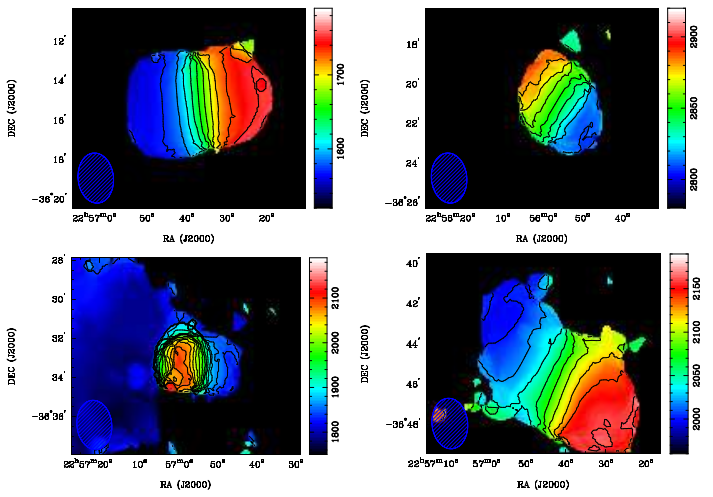} 
\caption{Low-resolution ATCA \HI\ velocity fields of the four spirals in the 
  vicinity of the giant elliptical galaxy IC\,1459. 
  {\bf (Top Left:)} IC\,5269B (1550 -- 1770\kms, 20\kms). {\bf (Top Right:)} 
  IC\,5269A (2810 -- 2880\kms, 10\kms). {\bf (Bottom Left:)} IC\,5264 (1840 -- 
  2100\kms, 20\kms). {\bf (Bottom Right:)} NGC~7418A (2000 -- 2160\kms, 
  20\kms). Contour level ranges and step sizes are given in brackets. The 
  synthesised beam ($151\farcs9 \times 108\farcs7$) is shown in the bottom 
  left corner of each panel. These maps are not primary-beam corrected.}
\label{fig:velo}
\end{figure*}

\subsection{\HI\ in IC\,1459 group members} 

The ATCA \HI\ properties of the four spirals near IC\,1459, as obtained from 
our low angular resolution \HI\ data cubes (after primary beam correction), 
are summarised in Table~\ref{tab:HIgalaxies}. Due to the location of three 
galaxies near the 50\% sensitivity of the ATCA primary beam (see 
Fig.~\ref{fig:ATCAfield}), the quoted \HI\ fluxes are somewhat uncertain. 

The \HI\ structure and velocity fields of IC\,5269A and IC\,5269B, both 
located north-west of IC\,1459, appear regular and symmetric, while NGC~7418A 
and IC~5264 show mildly distorted \HI\ distributions (see Fig.~\ref{fig:velo}).
High-resolution ASKAP \HI\ maps of the galaxies, which are discussed below,
are also shown by \citet{2015MNRAS.452.2680S}.

\begin{table} 
\caption{ATCA \HI\ properties of the detected IC\,1459 group members.}
\begin{tabular}{lccccc}
\hline
\hline
Galaxy & velocity range & \FHI      & \MHI \\
       &  [\kms]       & [Jy\kms] & [$10^8$\Msun]  \\
\hline
 IC\,5269B & 1550 -- 1780 &  18  $\pm$ 2.0 & 40  \\
 IC\,5269A & 2770 -- 2920 &  1.6 $\pm$ 0.4 & 3.5 \\
 IC\,5264  & 1800 -- 2150 &  0.5 $\pm$ 0.2 & 1.2 \\
 NGC~7418A & 2000 -- 2200 &  14  $\pm$ 1.3 & 31  \\
\hline
\label{tab:HIgalaxies}
\end{tabular}
\end{table}

\begin{itemize}

\item IC\,5269B (HIPASS J2256--36b) is a late-type spiral galaxy located 
    14\farcm5 north-west of IC\,1459 and has a similar system velocity. 
    \citet{2007MNRAS.378..137S} and \citet{2015MNRAS.452.2680S} show 
    high-resolution GMRT and ASKAP \HI\ maps of IC\,5269B, respectively, 
    indicating a symmetric and regularly rotating disc. We measure \FHI\ of 
    at least 18 Jy\kms, somewhat less than detected by ASKAP and Parkes 
    \citep{2015MNRAS.452.2680S}, and derive \MHI\ = $4 \times 10^9$\Msun.

\item IC\,5269A (HIPASS J2255--36) is a Magellanic barred spiral galaxy 
    located 16\farcm5 north-west of IC\,1459. Its nearly face-on
    stellar disc appears very irregular \citep{2007ApJS..173..404B}, while 
    the \HI\ data indicate a regularly rotating disc. 
   This galaxy is in the field of view, but unlikely a member of the group.
     Its systemic velocity 
    of 2845\kms\ is more than 1000\kms\ higher than that of IC\,1459. 
    We derive \MHI\ = $3.5 \times 10^8$\Msun.
   
\item IC\,5264 (\vopt\ = 2048\kms) is a small edge-on spiral galaxy, located 
    only 6\farcm5 south-west of IC\,1459 (\vopt\ = 1692\kms) and just west of 
    the \HI\ debris complex (\vHI\ $\sim$ 1800\kms). A partial dust lane gives 
    it a slightly peculiar appearance in optical images \citep{1995AJ....109.1576F}. IC\,5264's optical velocity is $\sim$350\kms\ higher than that of IC\,1459, either suggesting a larger distance and/or significant peculiar motions. The ATCA $pv$-diagram (Fig.~\ref{fig:HI-pv}) shows IC\,5264's \HI\ emission to cover a velocity range from $\sim$1800 to 2150\kms; the observed gradient is a clear indicator of galaxy rotation. The blue-shifted \HI\ emission of IC\,5264 on its western side (see Fig.~\ref{fig:velo}) falls within the velocity range of the extended \HI\ debris complex to the east. We estimate its \HI\ mass as $1.2 \times 10^8$\Msun. We also identified an \HI\ cloud in the direction of IC\,5264, at a very different radial velocity (C6 at \vsys\ $\sim 1500$\kms). 

\begin{figure} 
 \includegraphics[width=0.5\textwidth]{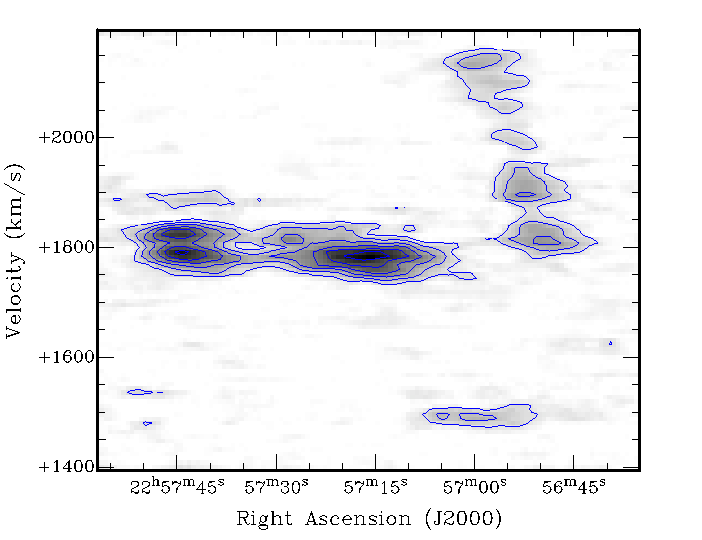} 
\caption{ATCA \HI\ position-velocity ($pv$) diagram of the area near IC~1459 
  shown in Figure~\ref{fig:HIclouds}. It includes the \HI\ cloud complex (1750 -- 1840\kms),
  the galaxy IC\,5264 ($\sim$1800 -- 2150\kms), and the \HI\ cloud C6 
  ($\sim$1500\kms). We applied 3-point Hanning smoothing to the velocities. 
  The contour levels are 3, 6 and 10 mJy\,beam$^{-1}$.}
\label{fig:HI-pv}
\end{figure}

\item NGC~7418A (HIPASS J2256--36a) is a late-type spiral galaxy, which lies  
    19\farcm5 south of IC\,1459. GALEX images show an extended ultraviolet
    (XUV) disc and \citet{2007ApJS..173..538T} note its two armed 
    feathery-looking spiral pattern of UV-bright clumps. NGC~7418A's ATCA 
    \HI\ distribution extends a factor three beyond the XUV disc and reveals 
    a tail oriented north-east, towards the main \HI\ debris complex 
    (Figs.~\ref{fig:ATCAfield} \& \ref{fig:velo}). We measure \FHI\ = 14 
    Jy\kms, corresponding to \MHI\ = $3.1 \times 10^9$\Msun. Furthermore, 
    we find an \HI\ cloud (C7, $\sim$1700\kms) just east of NGC~7418A, 
    offset from the systemic velocity, with an \HI\ mass of $1.5 \times 
    10^8$\Msun.

\end{itemize}

\section{Discussion}

Here we discuss the inner region of the IC\,1459 loose group where \HI\ debris 
extending over $\sim$250 kpc with a total \HI\ mass of $\sim$10$^9$\Msun\ have 
been detected with the ATCA. The presence of this unsettled gas strongly 
indicates interactions between the group members within the group potential.

\HI\ debris of such irregular morphology have been observed in numerous galaxy
systems. For example, a $\sim$300~kpc long \HI\ tail consisting of numerous
clouds was detected near the Hickson compact group 44 \citep{2013MNRAS.428..370S} with 
an \HI\ mass of at least $5 \times 10^8$\Msun. A huge \HI\ debris complex, 
known as the ``Vela Cloud'', with at least $3 \times 10^9$\Msun\ was identified 
by \citet{2010AJ....139..102E} near NGC~3263 in the NGC~3256 group. Extended
\HI\ filaments/debris are also observed in the Ursa Major cluster \citep{2012MNRAS.422.1835S} near the lenticular galaxies NGC~4026 and NGC~4011. 

Hydrodyamical simulations by \citet{2005MNRAS.357L..21B} demonstrate that the
group tidal field can be responsible for stripping of \HI\ from the outer disc
of gas-rich galaxies, while having little effect on their stellar discs. This
scenario was explored in detail to explain the origin of the massive \HI\ cloud
(HIPASS J0731--69) discovered near NGC~2442 by \citet{2001ApJ...555..232R}. Gas
stripping is found to be most efficient at the pericenter of the orbit on first
approach. In our case, it is possible that the galaxy IC\,5264 was stripped of
its outer \HI\ disc during infall to IC\,1459. Using the scaling relations by 
\citet{2014MNRAS.444..667D}
we find IC\,5264 to be \HI-deficient, which supports this
scenario. Furthermore, the ASKAP \HI\ maps presented by \citet{2015MNRAS.452.2680S}
show the \HI\ gas in IC\,5264 to be offset from its stellar disc towards the 
south-east. Tidal interactions between NGC~7418A and IC\,1459 are also likely,
supported by the north-eastern \HI\ tail of NGC~7418A, which links to the \HI\
debris complex near IC\,1459. Their projected separation of 19\farcm5 
corresponds to 165 kpc at the adopted group distance of 29 Mpc. The difference
in their systemic velocities ($\sim$400\kms) is likely due to peculiar motions 
within the group. The existence of intragroup X-ray emission as well as the 
peculiar features of IC\,1459 (e.g., stellar shells and a counter-rotating 
ring) further support the suggested interaction scenario. 

The detection of an \HI\ absorption line at $\sim$1714\kms\ towards
PKS~2254--367 implies the presence of cold hydrogen clumps in front of 
IC\,1459, likely part of the debris complex detected in \HI\ emission. This 
is further supported by the detection of CO(1--0) emission at 1782\kms\ 
towards IC\,1459 by \citet{2001A&A...374..421B}. 

Furthermore, we note that IC\,5269B and IC\,1459, which are separated by 
14\farcm5 (122 kpc), are the only two large galaxies in the group with 
systemic velocities in the velocity range of the \HI\ debris. No signs of
tidal interaction are visible in IC\,5269B. The \HI\ clouds detected near
IC\,5270 and NGC~7418 \citep{2015MNRAS.452.2680S}, both members of the IC\,1459 
group but outside the area studied here, highlight the complexity of the 
on-going gas stripping processes on small and large scales.

\section{Summary \& Outlook} 

Using Australia Telescope Compact Array 21-cm imaging we find a large \HI\ 
debris complex near the elliptical galaxy IC\,1459. The detected \HI\ clouds
span $\sim$250 kpc and have a total \HI\ mass of $\sim$10$^9$\Msun. Most of
the \HI\ debris are found to the east of the small edge-on spiral IC\,5264. 

Based on simulations by \citet{2005MNRAS.357L..21B} we suggest that the 
IC\,1459 group tidal field is responsible for stripping of \HI\ from the outer 
disc of IC\,5264, which is notably \HI-deficient, and NGC~7418A during their
first infall to the group. Enhanced star formation on NGC~5264's western side,
evident in GALEX $UV$ images, may have been caused by the tidal interactions.
No signs of star formation activity are detected in any of the identified 
\HI\ density enhancements. Deep imaging of the molecular gas distribution 
within the \HI\ debris complex would allow us to pin-point likely locations 
of star formation.  \\

The 21-cm spectral line of neutral hydrogen is an excellent tracer of galaxy 
interactions, providing spatial as well as kinematic information of the \HI\ 
gas in and between galaxies. It also allows us to derive both the \HI\ and 
total mass distributions of galaxies and groups, pin-point the locations of 
likely star formation in the galaxy outskirts and tidal dwarf galaxies in
the densest \HI\ debris.  
Although the phenomena related to galaxy interactions in groups and clusters 
are expected to be relatively common
\citep[e.g.,][]{2001A&A...377..812V,2005MNRAS.357L..21B,
2005MNRAS.363L..21B,2009AJ....138.1741C,2010AJ....139..102E,2012MNRAS.422.1835S,2013MNRAS.428.1790W,
2015MNRAS.452.2680S}
the need for high sensitivity
to extended structures as well as suitable velocity and angular resolution 
make them difficult to detect and study in detail. Future \HI\ observations 
with SKA Pathfinders such as ASKAP, Apertif and MeerKat \citep[see][]{2015IAUS..309...39K} 
will allow deep \HI\ imaging of many galaxy groups and clusters near and far. 
Building on the first wide-field \HI\ results from ASKAP \citep{2015MNRAS.452.2680S},
we expect the discovery of many \HI\ filaments, plumes, and clouds near 
interacting galaxies, providing new and valuable information on the evolution 
of galaxy groups.

\section*{Acknowledgements}

We thank the anonymous referee for the constructive comments, which very much
helped to improve the paper.
The observations presented here were obtained with the Australia Telescope 
Compact Array (ATCA) and the 64-m Parkes telescope which are funded by the 
Commonwealth of Australia for operation as a National Facility managed by 
CSIRO. This research has made use of the NASA/IPAC Extragalactic Database 
(NED) which is operated by the Jet Propulsion Laboratory, California Institute 
of Technology, under contract with the National Aeronautics and Space 
Administration and of SIMBAD database, operated at CDS, Strasbourg, France.


\bibliographystyle{mnras}
\bibliography{js-biblio} 

\end{document}